\documentclass[preprint,showpacs,prb,amsmath,amssymb]{revtex4}

\usepackage{graphicx}% Include figure files
\usepackage{dcolumn}% Align table columns on decimal point
\usepackage{bm}% bold math

\usepackage{natbib}
\renewcommand{\abstractname}{}

\renewcommand{\thefootnote}{\alph{footnote}}

\begin{document}

\title{A Family of One-Dimensional Vlasov-Maxwell Equilibria for the Force-Free Harris Sheet}
% \and 
\author{F. Wilson}
\email{fionaw@mcs.st-and.ac.uk}
\author{T.Neukirch}
\email{thomas@mcs.st-and.ac.uk}

\affiliation{School of Mathematics and Statistics, University of St. Andrews, 
St. Andrews KY16 9SS, United Kingdom}

\begin{abstract}
\noindent A family of self-consistent collisionless distribution functions for the force-free Harris sheet is presented. This family includes the distribution function recently found by Harrison and Neukirch [Phys. Rev. Lett. \textbf{102}, 135003 (2009)] as well as distribution functions with a different dependence on the particle energy, but with the same dependence on the canonical momenta. It is shown generally that the other distribution functions in the family give rise to the same pressure function and thus to the same current density and magnetic field as the known distribution function, provided certain conditions on the parameters are satisfied. A number of examples of distribution functions from the new family are given, which illustrate the use of the general method.
\end{abstract}
\pacs{52.20.-j, 52.25.Xz, 52.55.-s, 52.65.Ff}

\maketitle

\section{Introduction}
\noindent Investigations of plasma instabilities and plasma waves, both in astrophysics and in the laboratory, frequently start with a consideration of equilibrium solutions. For collisionless plasmas, the required equilibria are found by solving the steady-state Vlasov-Maxwell equations. Collisionless plasma equilbria have been investigated for many years, and a number of examples can be found in Refs.~\onlinecite{Grad-1961,Harris-1962,Nicholson-1963,Sestero-1964,Sestero-1967,Lam-1967,Kan-1972,Channell-1976,Bobrova-1979,Mynick-1979a,Greene-1993,Neukirch-1993,Attico-1999,Bobrova-2001,Schindler-2002,Fu-2005,Harrison-2009a,Harrison-2009b,Neukirch-2009}.\\
\indent Force-free magnetic fields, which satisfy $\textbf{j}\times\textbf{B}=\textbf{0}$, such that the current density and magnetic field are parallel to each other, can be used to model magnetic fields in low-beta plasmas, such as that of the solar corona.  Finding collisionless force-free equilibria is, however, not a trivial task, and hence there are few known examples. Of these known examples, only one is of the nonlinear force-free type \cite{Harrison-2009b,Neukirch-2009} (the rest are linear force-free \cite{Sestero-1967,Channell-1976,Bobrova-1979,Bobrova-2001}).\\
\indent It is well known that the steady-state Vlasov equation has many solutions. In fact, any positive function depending only on the constants of motion is a solution, provided the velocity moments exist. It can be shown that Amp\`{e}re's law, $\nabla\times\textbf{B}=\mu_0\textbf{j}$, can be written in terms of derivatives of the pressure function with respect to the components of the vector potential. An obvious consequence of this result is that two distribution functions giving rise to the same pressure function will automatically satisfy the Vlasov-Maxwell equations for the same magnetic field profile.\\
\indent In the present paper, it is shown how this property can be used to calculate a new family of distribution functions for the force-free Harris sheet. This family includes the previously known distribution function \cite{Harrison-2009b,Neukirch-2009}, as well as more general distribution functions with a different dependence on the particle energy, but with the same dependence on the canonical momenta. A direct comparison can be made between the pressure function calculated from the more general distribution functions and the pressure function calculated from the known distribution function in order to give conditions on the parameters which, when satisfied, will mean that the general distribution functions are also valid solutions for the force-free Harris sheet.\\
\indent The idea of finding new distribution functions for a given magnetic field profile by changing the dependence on the particle energy has been used before, but only for cases depending on the particle energy and a single component of the canonical momentum. Fu and Hau \cite{Fu-2005}, for example,  showed that kappa type distribution functions can be used for the Harris sheet magnetic field profile \cite{Harris-1962}. More recently, Kocharovsky et al. \cite{kocharovsky-2010} discussed distribution functions with an arbitrary dependence on the particle energy and a fixed dependence on one component of the canonical momentum for the relativistic case. It must, however, be emphasized that, for finding force-free Vlasov-Maxwell equilibria, it is crucial that the distribution functions depend on two components of the canonical momentum, and that the magnetic field has more than one non-zero component. \cite{Sestero-1967,Channell-1976,Bobrova-1979,Bobrova-2001,Harrison-2009a,Harrison-2009b,Neukirch-2009}.\\
\indent In the present paper, the general theory behind the work is given in Section \ref{sec:general_theory}, then the force-free Harris sheet model is discussed in Section \ref{sec:ffhs}. Section \ref{sec:method} contains a discussion of the method used to find the new equilibria, and this method is illustrated in Section \ref{sec:examples}, where three examples are given. Finally, Section \ref{sec:summary} contains a summary and conclusions.

\section{General Theory}

\label{sec:general_theory}

\noindent Finding Vlasov-Maxwell equilibria involves solving the steady-state Vlasov equation, 

\begin{equation}
 \textbf{v}\cdot\frac{\partial f_s}{\partial \textbf{r}} + \frac{q_s}{m_s}(\textbf{E} + \textbf{v}\times{\textbf{B}})\cdot\frac{\partial f_s}{\partial \textbf{v}}=0,
\label{ssv}
\end{equation}
together with the steady-state Maxwell equations,
%\begin{subequations}
\begin{eqnarray}
\nabla\cdot\textbf{E}&=&\frac{\sigma}{\epsilon_0},\label{max1}\\
\nabla\times\textbf{E}&=&0,\label{max2}\\
\nabla\times\textbf{B}&=&\mu_0\textbf{j},\label{max3}\\
\nabla\cdot\textbf{B}&=&0.\label{max4}
\end{eqnarray}
%\end{subequations}
In the present model, it is assumed that each quantity varies only with the $z$-coordinate, so that there is spatial invariance with respect to the $x$- and $y$-coordinates, and also time independence (since the system is in equilibrium). Furthermore, it is assumed that the magnetic field vanishes in the $z$-direction, and that it can be written as $\textbf{B}=\nabla\times{\textbf{A}}$, where $\textbf{A}=(A_x,A_y,0)$ is a vector potential whose $z$-component also vanishes. The $x$- and $y$-components of \textbf{B} are then given by
%\begin{subequations}
\begin{eqnarray}
 B_x&=&-\frac{dA_y}{dz},\label{bx}\\
B_y&=&\frac{dA_x}{dz},
\label{by}
\end{eqnarray}
%\end{subequations}
and so Eq. (\ref{max4}) is clearly satisfied. It is also assumed that the electric field \textbf{E} can be written as $\textbf{E}=-\nabla\phi$, where $\phi$ is an electric potential, so that
\begin{equation}
 E_z=-\frac{d\phi}{dz},\nonumber
\end{equation}
and so the electric field clearly satisfies Eq. (\ref{max2}). A further assumption is that the plasma consists of two particle species (ions and electrons) with charges $q_i=e$ and $q_e=-e$.

% that the plasma consists of two particle species, ions and electrons, and, secondly, that strict charge neutrality holds, which corresponds to the conditions $\phi=0$ and $\sigma=0$, where $\sigma$ is the charge density. The second assumption is valid if $L$, the typical length scale of the plasma (which represents the thickness of the current sheet in the present model), is much larger than the Debye length, $\lambda_D$. %In general, the assumption of quasineutrality does not imply that the electric field vanishes.

Due to the symmetries of the system, there are three constants of motion, namely the Hamiltonian (particle energy), $H_s$, arising from the time independence of the system, given by
\begin{equation}
 H_s=\frac{1}{2}m_s(v_x^{2}+v_y^{2}+v_z^{2})+q_s\phi,\nonumber
\end{equation}
the canonical momentum in the $x$-direction, $p_{xs}$, arising from the spatial symmetry in the $x$-direction, given by
\begin{equation}
 p_{xs}=m_{s}v_{x}+q_{s}A_{x},\nonumber
\end{equation}
and the canonical momentum in the $y$-direction, $p_{ys}$, arising from the spatial symmetry in the $y$-direction, given by
\begin{equation}
 p_{ys}=m_{s}v_{y}+q_{s}A_{y},\nonumber
\end{equation}
where $m_s$ is the mass of particle species $s$.\\ 
\indent Solving Eq. (\ref{ssv}) by the method of characteristics gives a set of functions $f_s=f_s(H_s,p_{xs},p_{ys})$, which depend only on the constants of motion. These are distribution functions for species $s$.
The remaining Maxwell equations to be solved are Gauss' law (Eq. (\ref{max1})) and Amp\`{e}re's law (Eq. (\ref{max3})), which can be expressed as
\begin{eqnarray}
-\epsilon_0\frac{d^2\phi}{dx^2}&=&\sigma,\nonumber\\
-\frac{1}{\mu_0}\frac{d^2A_x}{dx^2}&=&j_x,\nonumber\\
-\frac{1}{\mu_0}\frac{d^2A_y}{dx^2}&=&j_y,\nonumber
\end{eqnarray}
where the charge density $\sigma$ and the $x$- and $y$-components of the current density, $j_x$ and $j_y$, can be expressed as %velocity moments of the distribution functions as follows,
\begin{eqnarray}
\sigma(A_x,A_y,\phi)&=&\sum_sq_s\int_{-\infty}^{\infty}f_s(H_s,p_{xs},p_{ys})d^3v=-\frac{\partial{P_{zz}}}{\partial\phi},\label{sigma_def}\\
j_x(A_x,A_y,\phi)&=&\sum_sq_s\int_{-\infty}^{\infty}v_xf_s(H_s,p_{xs},p_{ys})d^3v=\frac{\partial{P_{zz}}}{\partial{A_x}},\label{ampere1}\\
j_y(A_x,A_y,\phi)&=&\sum_sq_s\int_{-\infty}^{\infty}v_yf_s(H_s,p_{xs},p_{ys})d^3v=\frac{\partial{P_{zz}}}{\partial{A_y}},\label{ampere2}
\end{eqnarray}
where $P_{zz}$ is the $zz$-component of the pressure tensor, defined as
\begin{equation}
P_{zz}(A_x,A_y,\phi)=\sum_sm_s\int{v_z^2f_s}d^3v.
\label{pz}
\end{equation}
\\
\indent Throughout this paper, strict charge neutrality ($\phi=0$) will be assumed, in line with previous work \cite{Channell-1976,Harrison-2009a,Harrison-2009b,Neukirch-2009}, leading to the condition $\sigma=0$. Ampere's Law can then be written as \cite{Harrison-2009a}
\begin{eqnarray}
\frac{d^2A_x}{dz^2}=-\mu_0\frac{\partial{P_{zz}}}{\partial{A_x}},\label{ampere3}\\
\frac{d^2A_y}{dz^2}=-\mu_0\frac{\partial{P_{zz}}}{\partial{A_y}}.\label{ampere4}
\end{eqnarray}
If the distribution functions are known, it is relatively straightforward to calculate the magnetic field profile by solving Eqs. (\ref{ampere3}) and (\ref{ampere4}). However, determining a distribution function for a given magnetic field profile is much more difficult and involves solving Eq. (\ref{pz}) as an integral equation \cite{Channell-1976,Mynick-1979a}. Distribution functions giving rise to the same pressure function $P_{zz}(A_x,A_y)$ will give rise to the same current density through Eqs. (\ref{ampere1}) and (\ref{ampere2}), and hence the same magnetic field profile through Eqs. (\ref{ampere3}) and (\ref{ampere4}). This idea is central to the method employed in Section \ref{sec:method} to find new equilibrium distribution functions for the force-free Harris sheet.

\section{The Force-Free Harris Sheet}

\label{sec:ffhs}

\begin{figure}[htp]
\centering
\scalebox{0.6}{\includegraphics{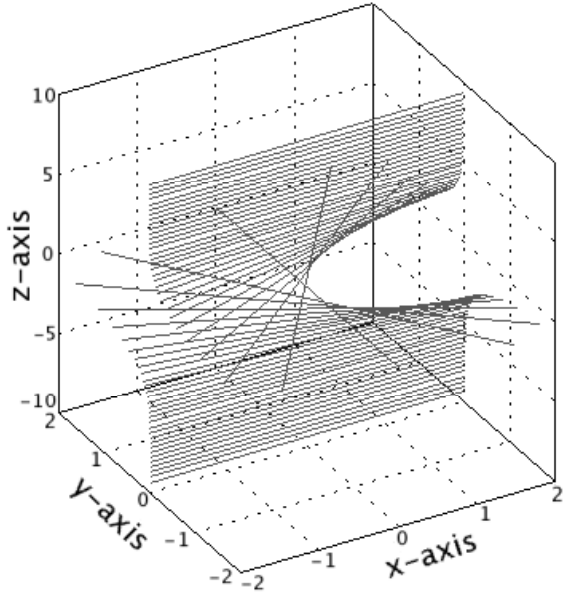}}
\caption{Field line plot for the force-free Harris sheet.}\label{ffharris_fieldlines}
\end{figure}

\begin{figure}[htp]
\centering
\scalebox{0.3}{\includegraphics{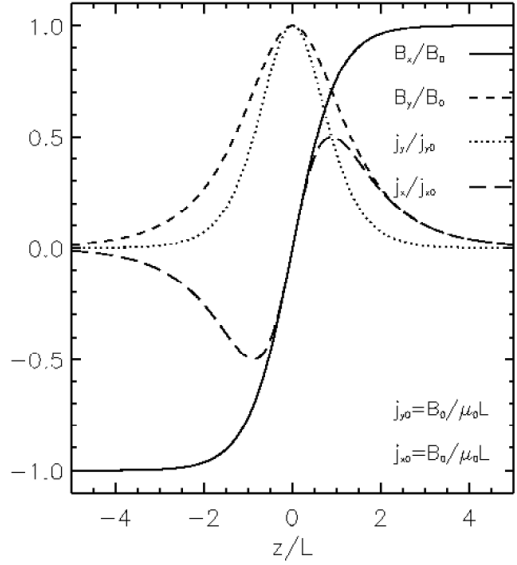}}
\caption{Plot of the magnetic field and current density profiles for the force-free Harris sheet.}\label{ffharris}
\end{figure}

\noindent The force-free Harris sheet has a magnetic field profile as follows,
\begin{eqnarray}
B_{x,ffhs}&=&B_0\tanh(z/L),\nonumber\\
B_{y,ffhs}&=&\frac{B_0}{\cosh(z/L)},\nonumber
\end{eqnarray}
where $B_0$ is a constant and $L$ is a parameter which specifies the thickness of the sheet (as mentioned in Section \ref{sec:general_theory}). The $x$-component of the field is the same as that of the Harris sheet \cite{Harris-1962}, and the addition of the $y$-component makes the field force-free, since $B_x^2+B_y^2=B_0^2$. The $y$-component of the field also maintains the force balance across the sheet, since both the plasma pressure $P_{zz}$ and the magnetic pressure $B^2/2\mu_0=(B_x^2+B_y^2)/2\mu_0$ are constant with respect to $z$ \cite{Harrison-2009a,Harrison-2009b,Neukirch-2009}, but the magnetic field components vary with $z$. Fig. \ref{ffharris_fieldlines} shows a plot of the field lines for the force-free Harris sheet.
Using Eqs. (\ref{bx}) and (\ref{by}) gives the components of the vector potential as
\begin{eqnarray}
 A_{x,ffhs}&=&2B_0L\tan^{-1}(e^{z/L}),\label{ax}\nonumber\\
A_{y,ffhs}&=&-B_0L\ln\left[\cosh(z/L)\right]\label{ay}.\nonumber
\end{eqnarray}
The current density can be calculated from Amp\`{e}re's Law, and is given by
\begin{eqnarray}
j_{x,ffhs}&=&\frac{B_0}{\mu_0L}\frac{\tanh(z/L)}{\cosh(z/L)},\nonumber\\
j_{y,ffhs}&=&\frac{B_0}{\mu_0L}\frac{1}{\cosh^2(z/L)}.\nonumber
\end{eqnarray}
Fig. \ref{ffharris} shows the magnetic field and current density profiles for the force-free Harris sheet.
The magnetic field is a nonlinear force-free field, and so the current density is parallel to the magnetic field, \begin{equation}
\textbf{j}_{ffhs}=\frac{\alpha(z)}{\mu_0}\textbf{B}_{ffhs},\nonumber
\end{equation}
with the force-free parameter $\alpha(z)$ given by
\begin{equation}
\alpha(z)=\frac{1}{L}\frac{1}{\cosh(z/L)},\nonumber
\label{alpha}
\end{equation}
which is constant along a given field line, but varies from field line to field line.

A distribution function satisfying the steady-state Vlasov-Maxwell equations for the force-free Harris sheet, which was found by Harrison and Neukirch \cite{Harrison-2009b}, is given by,
\begin{equation}
 f_{s,ffhs}=\frac{n_{0s}}{(\sqrt{2\pi}v_{th,s})^3}\exp(-\beta_sH_s)\left[a_s\cos(\beta_su_{xs}p_{xs})+\exp(\beta_su_{ys}p_{ys})+b_s\right],
\label{fs1}
\end{equation}
where $a_s,b_s,u_{xs},u_{ys},n_{0s}$ and $\beta_s$ are constant parameters of the distribution function, and $v_{th,s}$ is the thermal velocity of particle species $s$. Note that the condition $b_s>a_s$ must hold to ensure that $f_s$ is positive. The \textit{zz}-component of the pressure tensor, $P_{zz,ffhs}$, is given in terms of the microscopic notation by
\begin{equation}
 P_{zz,ffhs}(A_x,A_y)=\frac{\beta_e+\beta_i}{\beta_e\beta_i}n_0\left[\frac{1}{2}\cos(e\beta_eu_{xe}A_x)+\exp(-e\beta_eu_{ye}A_y)+b\right].
\label{p}
\end{equation}
A detailed discussion of the derivation of Eqs. (\ref{fs1}) and (\ref{p}) can be found in Ref.~\onlinecite{Neukirch-2009}.
In the next section, it is shown how to calculate a family of distribution functions for the force-free Harris sheet, which includes the distribution function (\ref{fs1}), as well as distribution functions with a different dependence on the particle energy $H_s$. 

\section{Finding a Family of Distribution Functions for the Force-Free Harris Sheet}

\label{sec:method}

\noindent Consider a more general distribution function of the form

\begin{equation}
f_{s,g}=\frac{m_s^3}{q_s^4}f_0h\left(\frac{m_sH_s}{q_s^2}\right)[a_s\cos(\beta_su_{xs}p_{xs})+\exp(\beta_su_{ys}p_{ys})+b_s],
\label{more_general_df}
\end{equation}
where $h$ is an arbitrary function of $(m_sH_s)/q_s^2$ and $f_0$ is a positive constant. The distribution function (\ref{more_general_df}) is of the same form as the distribution function given by Eq. (\ref{fs1}), but it has a different dependence on the particle energy. It can be shown that, when calculating velocity moments of distribution functions of the form
\begin{equation}
f_{s,g}=\frac{m_s^3}{q_s^4}F\left(\frac{m_sH_s}{q_s^2},\frac{p_{xs}}{q_s},\frac{p_{ys}}{q_s}\right),\label{fs_german}
\end{equation}
it is possible to make the resulting integrals independent of the particle species $s$ whenever $\phi=0$, that is, whenever strict charge neutrality is assumed. This has been shown before for distribution functions depending only on energy and one component of the canonical momentum \cite{Schmid-1965}, and can be used to show that distribution functions of the form (\ref{more_general_df}) are solutions of the Vlasov-Maxwell equations for the force-free Harris sheet, provided certain conditions on the parameters are satisfied.

The $zz$-component of the pressure tensor, $P_{zz,g}$, can be written as
\begin{eqnarray}
P_{zz,g}(A_x,A_y,\phi)=\sum_s\frac{1}{m_s^4}\int{p_{zs}^2}f_{s,g}d^3p,\nonumber
\end{eqnarray}
where $d^3p=dp_{xs}dp_{ys}dp_{zs}$. At this point, the dependence of $P_{zz,g}$ on $\phi$ is still stated explicitly. In the later parts of the paper, however, the assumption $\phi=0$ will be made. \\
\indent Defining three new variables $E$, $P$ and $Q$ as
\begin{eqnarray}
E&=&\frac{m_sH_s}{q_s^2},\nonumber\\
P&=&\frac{p_{xs}}{q_s},\nonumber\\
Q&=&\frac{p_{ys}}{q_s},\nonumber
\end{eqnarray}
and using these as integration variables, gives the pressure as
\begin{eqnarray}
P_{zz,g}(A_x,A_y,\phi)&=&2f_0\sum_s\frac{|q_s|}{m_s}\int_{-\infty}^{\infty}\int_{-\infty}^{\infty}\int_{E_{min}}^{\infty}h(E)[a_s\cos(\beta_su_{xs}q_sP)+\exp(\beta_su_{ys}q_sQ)+b_s]\nonumber\\
&{}&\times\left[2E-\frac{2m_s\phi}{q_s}-(P-A_x)^2-(Q-A_y)^2\right]^{1/2}dEdPdQ,
\label{general_pzz}
\end{eqnarray}
where
\begin{equation}
E_{min}=\frac{1}{2}[(P-A_x)^2+(Q-A_y)^2]+\frac{m_s\phi}{q_s}\nonumber.%\label{e_min}.
\end{equation}
The charge density, $\sigma_g$, can be calculated from $P_{zz,g}$ using the definition in Eq. (\ref{sigma_def}), and is given by
\begin{eqnarray}
\sigma_g&=&2f_0\sum_s\frac{|q_s|}{q_s}\int_{-\infty}^{\infty}\int_{-\infty}^{\infty}\int_{E_{min}}^{\infty}h(E)[a_s\cos(\beta_su_{xs}q_sP)+\exp(\beta_su_{ys}q_sQ)+b_s]\nonumber\\
&{}&\times\left[2E-\frac{2m_s\phi}{q_s}-(P-A_x)^2-(Q-A_y)^2\right]^{-1/2}dEdPdQ.
\label{charge_d}
\end{eqnarray}
It can be seen that the terms inside the integration in Eq. (\ref{charge_d}) will all be independent of $s$, that is they will not depend upon the particle species, if the following three conditions are satisfied,
\begin{eqnarray}
\phi&=&0,\label{ind1}\\
e\beta_e|u_{xe}|&=&e\beta_i|u_{xi}|=\alpha,\label{ind2}\\
-e\beta_eu_{ye}&=&e\beta_iu_{yi}=\gamma.\label{ind3}
\end{eqnarray}
The above conditions are consistent with those obtained in Ref.~\onlinecite{Neukirch-2009}. Due to the modulus signs in Eq. (\ref{ind2}), the parameter $\alpha$ is always positive, and so $u_{xe}$ and $u_{xi}$ from the distribution function (\ref{more_general_df}) can be positive or negative, and can have the same or opposite sign from each other. The neutrality condition $\sigma=0$ gives
\begin{equation}
\sum_s\frac{|q_s|}{q_s}(a_sI_1+I_2+b_sI_3)=0,\nonumber
\end{equation}
where
\begin{eqnarray}
I_1&=&\int_{-\infty}^{\infty}\int_{-\infty}^{\infty}\int_{E_{min,0}}^{\infty}h(E)\cos(\alpha{P})\nonumber\\
&{}&\times\left[2E-(P-A_x)^2-(Q-A_y)^2\right]^{-1/2}dEdPdQ,\nonumber\\
I_2&=&\int_{-\infty}^{\infty}\int_{-\infty}^{\infty}\int_{E_{min,0}}^{\infty}h(E)\exp(\gamma{Q})\nonumber\\
&{}&\times\left[2E-(P-A_x)^2-(Q-A_y)^2\right]^{-1/2}dEdPdQ,\nonumber\\
I_3&=&\int_{-\infty}^{\infty}\int_{-\infty}^{\infty}\int_{E_{min,0}}^{\infty}h(E)\nonumber\\
&{}&\times\left[2E-(P-A_x)^2-(Q-A_y)^2\right]^{-1/2}dEdPdQ,\nonumber
\end{eqnarray}
where $E_{min,0}$ is the value of $E_{min}$ when $\phi=0$. This then gives
\begin{equation}
(a_i-a_e)I_1+(b_i-b_e)I_3=0,\nonumber
\end{equation}
which will of course be satisified if
\begin{eqnarray}
 a_e&=&a_i=A,\label{ind4}\\
b_e&=&b_i=B.\label{ind5}
\end{eqnarray}
Note that the condition $b_s>a_s$ mentioned in Section \ref{sec:ffhs} leads to the condition $B>A$, which must be satisfied to ensure that the distribution functions are positive. When the conditions (\ref{ind1})-(\ref{ind3}), (\ref{ind4}) and (\ref{ind5}) are satisifed, the pressure $P_{zz.g}$, given by Eq. (\ref{general_pzz}), can be rewritten as
\begin{eqnarray}
P_{zz,g}(A_x,A_y)=2f_0\frac{(m_e+m_i)e}{m_em_i}[AH_1\cos(\alpha{A_x})+H_2\exp(\gamma{A_y})+BH_3],
\label{yyy}
\end{eqnarray}
where
\begin{eqnarray}
H_1&=&\int_{-\infty}^{\infty}\int_{-\infty}^{\infty}\int_{E_{min,0}}^{\infty}h(E)\left[2E-S^2-T^2\right]^{1/2}\cos(\alpha{S})dEdSdT,\nonumber\\
H_2&=&\int_{-\infty}^{\infty}\int_{-\infty}^{\infty}\int_{E_{min,0}}^{\infty}h(E)\left[2E-S^2-T^2\right]^{1/2}\exp(\gamma{T})dEdSdT,\nonumber\\
H_3&=&\int_{-\infty}^{\infty}\int_{-\infty}^{\infty}\int_{E_{min,0}}^{\infty}h(E)\left[2E-S^2-T^2\right]^{1/2}dEdSdT,\nonumber
\end{eqnarray}
with $S=P-A_x$ and $T=Q-A_y$. This can be compared with the pressure (\ref{p}), and so the general pressure given by Eq. (\ref{yyy}) will be equal to the pressure (\ref{p}) if the following additional conditions are satisfied,
\begin{eqnarray}
\frac{2(m_e+m_i)e}{m_em_i}f_0H_2&=&\frac{\beta_e+\beta_i}{\beta_e\beta_i}n_0,\label{hhh}\\
\frac{AH_1}{H_2}&=&\frac{1}{2},\label{hhh1}\\
\frac{BH_3}{H_2}&=&b.\label{hhh2}
\end{eqnarray}
When these conditions are satisfied, in addition to conditions (\ref{ind1})-(\ref{ind3}), (\ref{ind4}) and (\ref{ind5}), Amp\`{e}re's law in the form given by Eqs. (\ref{ampere3}) and (\ref{ampere4}) is satisfied for the set of general distribution functions (\ref{more_general_df}) and, therefore, they form a family of equilibrium solutions of the Vlasov-Maxwell equations for the force-free Harris sheet, in addition to the known distribution function \cite{Harrison-2009b}. In the next section, three explicit examples are given, which show possible choices of the function $h(m_sH_s/q_s^2)=h(E)$. The validity of the conditions (\ref{hhh})-(\ref{hhh2}) will of course depend on the choice of the function $h(E)$, and this will be discussed in each of the three examples.

\section{Examples of New Distribution Functions}

\noindent The following three examples in this section illustrate the use of the method discussed in Section \ref{sec:method}, for various choices of the function $h(m_sH_s/q_s^2)=h(E)$.

\label{sec:examples}

\subsection{Delta Function}
\begin{figure}[htp]
\centering
\scalebox{0.5}{\includegraphics{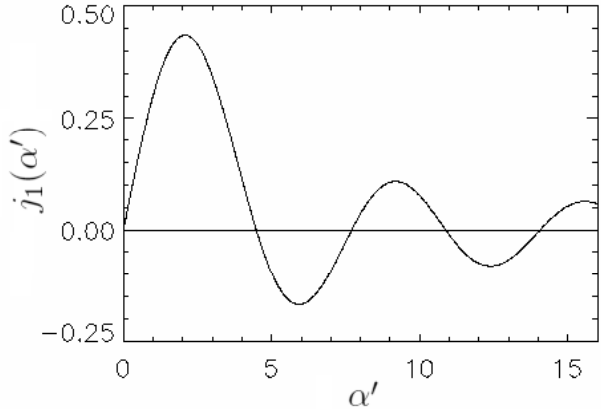}}
\caption{The spherical Bessel function $j_1(\alpha')$}\label{sph_1}
\end{figure}

\noindent Consider a distribution function of the form (\ref{more_general_df}), with the function $h(m_sH_s/q_s^2)=h(E)$ given by a delta function,
\begin{equation}
h(E)=\delta(E-E_0),\label{delta_df}
\end{equation}
where $E_0>E_{min,0}$. This corresponds to a case where the distribution function is zero everywhere except for one particular value of the energy ($E=E_0$), and so all particles are assumed to have the same energy. Carrying out the $E$-integration first gives $P_{zz,g}$ as
\begin{eqnarray}
P_{zz,g}(A_x,A_y)&=&2f_0\sum_s\frac{|q_s|}{m_s}\Bigg[A\cos(\alpha{A_x})\int_{-\infty}^{\infty}\int_{-\infty}^{\infty}(2E_0-S^2-T^2)^{1/2}\cos(\alpha{S})dSdT\nonumber\\
&{}&+\exp(\gamma{A_y})\int_{-\infty}^{\infty}\int_{-\infty}^{\infty}(2E_0-S^2-T^2)^{1/2}\exp(\gamma{T})dSdT\nonumber\\
&{}&+B\int_{-\infty}^{\infty}\int_{-\infty}^{\infty}(2E_0-S^2-T^2)^{1/2}dSdT\Bigg].\nonumber
\end{eqnarray}
Using cylindrical coordinates ($r$,$\theta$), with 
\begin{eqnarray}
S^2+T^2&=&2E_0r^2,\label{cyl1}\\
 S&=&\sqrt{2E_0}r\cos\theta\label{cyl2},\\
T&=&\sqrt{2E_0}r\sin\theta\label{cyl3},
\end{eqnarray}
then gives
\begin{eqnarray}
P_{zz,g}(A_x,A_y)&=&2\sqrt{8E_0^3}f_0\frac{(m_e+m_i)e}{m_em_i}\Bigg[A\cos(\alpha{A_x})\int_{0}^{2\pi}\int_{0}^{1}r(1-r^2)^{1/2}\cos(\alpha'{r\cos\theta})drd\theta\nonumber\\
&{}&+\exp(\gamma{A_y})\int_{0}^{2\pi}\int_{0}^{1}r(1-r^2)^{1/2}\exp(\gamma'{r\sin\theta})drd\theta+\frac{2\pi{B}}{3}\Bigg],\nonumber
\end{eqnarray}
where
\begin{eqnarray}
\alpha'=\sqrt{2E_0}\alpha,\label{a_prime}\\
\gamma'=\sqrt{2E_0}\gamma.\label{b_prime}
\end{eqnarray}
%The $\theta$-integrations can be carried out by using the formulae (\ref{theta2}) and (\ref{theta1}), and then the $r$-integrations can be carried out by using the formulae (\ref{app1}) and (\ref{app3}).
The remaining integrations can then be carried out by using the formulae (\ref{theta2})-(\ref{app3}) given in the Appendix. This gives $P_{zz,g}$ as
\begin{eqnarray}
P_{zz,g}(A_x,A_y)&=&4\pi\sqrt{8E_0^3}f_0\frac{(m_e+m_i)e}{m_em_i}\Bigg[A\sqrt{\frac{\pi}{2\alpha'^3}}J_{3/2}(\alpha')\cos(\alpha{A_x})\nonumber\\
&{}&+\sqrt{\frac{\pi{i}}{2\gamma'^3}}J_{3/2}(i\gamma')\exp(\gamma{A_y})+\frac{B}{3}\Bigg].
\label{pzz1}
\end{eqnarray}
The Bessel functions $J_{3/2}$ in Eq. (\ref{pzz1}), of fractional order, can be written in terms of spherical Bessel functions $j_1$ (of integer order) through the identity
\begin{equation}
j_n(z)=\sqrt{\frac{\pi}{2z}}J_{n+1/2}(z),
\label{spherical_bessel}
\end{equation}
which gives,
\begin{equation}
J_{3/2}(z)=\sqrt{\frac{2z}{\pi}}j_1(z),\nonumber
\end{equation}
where
\begin{equation}
j_1(z)=\frac{\sin{z}}{z^2}-\frac{\cos{z}}{z}
\label{sph_bess1}.
\end{equation}
This gives $P_{zz,g}$ as
%\begin{eqnarray}
%j_n(z)&=&\sqrt{\frac{\pi}{2z}}J_{n+\frac{1}{2}}(z),\label{bess}\\
%j_1(z)&=&\frac{\sin{z}}{z^2}-\frac{\cos{z}}{z}\label{bess2}
%\end{eqnarray}
%where $j_n$ is a spherical Bessel function of integer order $n$, and using the fact that $\Gamma(3/2)=\sqrt{\pi}/2$ gives
%\begin{equation}
%\int_{0}^{1}r(1-r^2)^{1/2}J_0(zr)dr=\frac{\sin{z}}{z^3}-\frac{\cos{z}}{z^2}.
%\end{equation}
%The second integral in equation (\ref{ggg}) can be evaluated in a similar way to (\ref{ggg1}), by using the definition
%\begin{equation}
%I_0(z)=J_0(iz),
%\end{equation}
%and is given by
%\begin{equation}
%\int_{0}^{1}r(1-r^2)^{1/2}I_0(xr)dr=\frac{\cosh{x}}{x^2}-\frac{\sinh{x}}{x^3},
%\end{equation}
\begin{eqnarray}
P_{zz,g}(A_x,A_y)&=&4\pi{f_0}\frac{(m_e+m_i)e}{m_em_i}\frac{\sqrt{8E_0^3}}{\gamma'^3}\left(\gamma'\cosh\gamma'-\sinh\gamma'\right)\nonumber\\
&{}&\times\Bigg[A\cos(\alpha{A_x})\frac{\gamma'^3}{\alpha'^3}\left(\frac{\sin\alpha'-\alpha'\cos\alpha'}{\gamma'\cosh\gamma'-\sinh\gamma'}\right)\nonumber\\
&{}&+\exp(\gamma{A_y})+\frac{B}{3}\left(\frac{\gamma'^3}{\gamma'\cosh\gamma'-\sinh\gamma'}\right)\Bigg].\label{delta_pzz}
\end{eqnarray}
The conditions (\ref{hhh})-(\ref{hhh2}) for the pressure (\ref{delta_pzz}) to be equal to the pressure (\ref{p}) are then given by
\begin{eqnarray}
\frac{\beta_e+\beta_i}{\beta_e\beta_i}n_0&=&4\pi{f_0}\frac{(m_e+m_i)e}{m_em_i}\frac{\sqrt{8E_0^3}}{\gamma'^3}\left(\gamma'\cosh\gamma'-\sinh\gamma'\right),\label{hhh3}\\
\frac{1}{2}&=&A\frac{\gamma'^3}{\alpha'^3}\left(\frac{\sin\alpha'-\alpha'\cos\alpha'}{\gamma'\cosh\gamma'-\sinh\gamma'}\right),\label{hhh4}\\
b&=&\frac{B}{3}\left(\frac{\gamma'^3}{\gamma'\cosh\gamma'-\sinh\gamma'}\right),\label{hhh5}
\end{eqnarray}
and the condition $B>A$ gives rise to the following condition on $b$
\begin{eqnarray}
b>\frac{1}{6}\left(\frac{\alpha'^3}{\sin\alpha'-\alpha'\cos\alpha'}\right).\label{hhh6}
\end{eqnarray}
The right-hand side of condition (\ref{hhh3}) is always positive since the $\gamma'$-dependent function, $(\gamma'\cosh\gamma'-\sinh\gamma')/\gamma'^3$, is always positive, regardless of the value of $\gamma'$. Note also that $m_s,e,E_0,f_0>0$. Condition (\ref{hhh3}) is, therefore, a valid condition since its left-hand side is also always positive ($\beta_s,n_0>0$). The constant $A$ in condition (\ref{hhh4}) is positive, and the $\gamma'$-dependent part is the same as in condition (\ref{hhh3}) which, as discussed, is always positive for any value of $\gamma'$. The $\alpha'$-dependent part is given by
\begin{equation}
\frac{\sin\alpha'-\alpha'\cos\alpha'}{\alpha'^3}=\frac{j_1(\alpha')}{\alpha'},
\end{equation}
which can be positive or negative due to the spherical Bessel function $j_1(\alpha')$, shown in Fig. \ref{sph_1}. Condition (\ref{hhh4}) is, therefore, only valid in the regions where $j_1(\alpha')>0$. It can be seen from Fig. \ref{sph_1} that it is valid for small values of $\alpha'$. Condition (\ref{hhh5}) is also a valid condition, since both the left- and right-hand sides are positive ($B$ is positive and the $\gamma'$-dependent part is positive). The fact that $\gamma'$ can be positive or negative means that the parameters $u_{ye}$ and $u_{yi}$ from the distribution function (\ref{more_general_df}) (with the function $h(m_sH_s/q_s^2)=h(E)$ given by Eq. (\ref{delta_df})) can be positive or negative, but must have opposite signs from each other through the definition (\ref{ind3}) of $\gamma=\gamma'/\sqrt{2E_0}$. 

\subsection{Step Function}\label{sec:step}

\noindent Consider a distribution function of the form (\ref{more_general_df}), with the function $h(m_sH_s/q_s^2)=h(E)$ given by a step function,
\begin{equation}
h(E)=\Theta(E_0-E)=\left\{ \begin{array}{ll}
1, & \textrm{$E\le{E_0}$}\\
0, & \textrm{$E_0<E$}
\end{array} \right..\label{step_df}
\end{equation}
The $E$-integral is the same for each part of $P_{zz,g}$, and is given by
\begin{eqnarray}
\int_{E_{min,0}}^{\infty}\Theta(E_0-E)(2E-S^2-T^2)^{1/2}dE&=&\int_{E_{min,0}}^{E_0}(2E-S^2-T^2)^{1/2}dE\nonumber\\
&=&\frac{1}{3}(2E_0-S^2-T^2)^{3/2}.\nonumber
\end{eqnarray}
Using a cylindrical coordinate system $(r,\theta)$ as in the first example (see Eqs. (\ref{cyl1})-(\ref{cyl3})), the resulting integrals can then be evaluated in a similar way to the previous example (see the Appendix for details). Then, using Eq. (\ref{spherical_bessel}) as in the previous example, the resulting Bessel functions (of order $5/2$) can be expressed in terms of the spherical Bessel function $j_2$, where
\begin{eqnarray}
j_2(z)&=&\left(\frac{3}{z^3}-\frac{1}{z}\right)\sin{z}-\frac{3}{z^2}\cos{z}.\nonumber
\end{eqnarray}
The pressure is then given by
\begin{eqnarray}
P_{zz,g}(A_x,A_y)&=&4\pi{f_0}\frac{(m_e+m_i)e}{m_em_i}\frac{\sqrt{32E_0^5}}{\gamma'^5}\left[(3+\gamma'^2)\sinh\gamma'-3\gamma'\cosh\gamma'\right]\nonumber\\
&{}&\times\Bigg[A\cos(\alpha{A_x})\frac{\gamma'^5}{\alpha'^5}\frac{(3-\alpha'^2)\sin\alpha'-3\alpha'\cos\alpha'}{(3+\gamma'^2)\sinh\gamma'-3\gamma'\cosh\gamma'}\nonumber\\
&{}&+\exp(\gamma{A_y})+\frac{B}{15}\frac{\gamma'^5}{(3+\gamma'^2)\sinh\gamma'-3\gamma'\cosh\gamma'}\Bigg],\label{step_pzz}
\end{eqnarray}
where $\alpha'$ and $\gamma'$ are given by Eqs. (\ref{a_prime}) and (\ref{b_prime}).
The conditions (\ref{hhh})-(\ref{hhh2}) for the pressure (\ref{step_pzz}) to be equal to (\ref{p}) are then given by
\begin{eqnarray}
\frac{\beta_e+\beta_i}{\beta_e\beta_i}n_0&=&4\pi{f_0}\frac{(m_e+m_i)e}{m_em_i}\frac{\sqrt{32E_0^5}}{\gamma'^5}\left[(3+\gamma'^2)\sinh\gamma'-3\gamma'\cosh\gamma'\right],\label{zzz1}\\
\frac{1}{2}&=&A\frac{\gamma'^5}{\alpha'^5}\frac{(3-\alpha'^2)\sin\alpha'-3\alpha'\cos\alpha'}{(3+\gamma'^2)\sinh\gamma'-3\gamma'\cosh\gamma'},\label{zzz2}\\
b&=&\frac{B}{15}\frac{\gamma'^5}{(3+\gamma'^2)\sinh\gamma'-3\gamma'\cosh\gamma'},\label{zzz3}
\end{eqnarray}
and the condition $B>A$ gives the following condition on $b$,
\begin{eqnarray}
b>\frac{1}{30}\frac{\alpha'^5}{(3-\alpha'^2)\sin\alpha'-3\alpha'\cos\alpha'}.
\end{eqnarray}
The right-hand side of condition (\ref{zzz1}) is always positive since the $\gamma'$-dependent function, $((3+\gamma'^2)\sinh\gamma'-3\gamma'\cosh\gamma')/\gamma'^5$, is always positive, regardless of the value of $\gamma'$. Note also that $m_s,e,E_0,f_0>0$. Condition (\ref{zzz1}) is, therefore, a valid condition since its left-hand side is also always positive ($\beta_s,n_0>0$). The constant $A$ in condition (\ref{zzz2}) is positive, and the $\gamma'$-dependent part is the same as in condition (\ref{zzz1}) which, as discussed, is always positive for any value of $\gamma'$. The $\alpha'$-dependent part is given by
\begin{equation}
\frac{(3-\alpha'^2)\sin\alpha'-3\alpha'\cos\alpha'}{\alpha'^5}=\frac{j_2(\alpha')}{\alpha'^2},
\end{equation}
which can be positive or negative due to the spherical Bessel function $j_2(\alpha')$, which has a similar profile (qualitatively) to that of the spherical Bessel function $j_1(\alpha')$, shown in Fig. \ref{sph_1}. Condition (\ref{zzz2}) is, therefore, only valid in the regions where $j_2(\alpha')>0$. Condition (\ref{zzz3}) is also a valid condition, since both the left- and right-hand sides are positive ($B$ is positive and the $\gamma'$-dependent part is positive). The fact that $\gamma'$ can be positive or negative means that the parameters $u_{ye}$ and $u_{yi}$ from the distribution function (\ref{more_general_df}) (with the function $h(m_sH_s/q_s^2)=h(E)$ given by (\ref{step_df})) can be positive or negative, but must have opposite signs from each other through the definition (\ref{ind3}) of $\gamma=\gamma'/\sqrt{2E_0}$.

\subsection{Power of $E_0-E$}
\noindent Consider a distribution function of the form (\ref{more_general_df}), with the function $h(m_sH_s/q_s^2)=h(E)$ given by
\begin{eqnarray}
h(E) = \left\{ \begin{array}{ll}
(E_0-E)^\chi, & \textrm{$E<E_0$}\\
0, & \textrm{$E_0\le{E}$}
\end{array} \right.,\label{chi}\nonumber
\end{eqnarray}
where $\chi>-1$. When calculating $P_{zz,g}$ using Eq. (\ref{yyy}), the $E$-integration is the same in each triple integral, and is given by
\begin{equation}
\int_{E_{min,0}}^{E_0}(E_0-E)^\chi[2(E-E_{min,0})]^{1/2}dE\label{psi2},
\end{equation}
which can be written as
\begin{equation}
\sqrt{2}\int_{0}^{\psi_0}\psi^\chi(\psi_0-\psi)^{1/2}d\psi\label{psi1},
\end{equation}
where $\psi=E_0-E$ and $\psi_0=E_0-E_{min,0}$. The integral (\ref{psi1}) can be evaluated by using the formula
\begin{equation}
\int_{0}^{z}t^{\nu-1}(z-t)^{\mu-1}dt=z^{\mu+\nu-1}B(\mu,\nu),\nonumber
\end{equation}
where $B(\mu,\nu)$ is a beta function defined by
\begin{equation}
B(\mu,\nu)=\frac{\Gamma(\mu)\Gamma(\nu)}{\Gamma(\mu+\nu)},\nonumber
\end{equation}
and $\Re\mu>0$, $\Re\nu>0$ (which gives $\chi>-1$). The integral (\ref{psi2}) is then given by
\begin{eqnarray}
\int_{E_{min,0}}^{E_0}(E_0-E)^\chi[2(E-E_{min,0})]^{1/2}dE\nonumber\\
=\frac{\sqrt{\pi}}{4.2^\chi}\frac{\Gamma(\chi+1)}{\Gamma+5/2}(2E_0-S^2-T^2)^{3/2+\chi},\nonumber
\end{eqnarray}
(note that $E_{min,0}=(1/2)(S^2+T^2)$). Using a cylindrical coordinate system as before (see Eqs. (\ref{cyl1})-(\ref{cyl3})) gives $P_{zz,g}$ as
\begin{eqnarray}
P_{zz,g}(A_x,A_y)&=&\frac{f_0(m_e+m_i)e}{m_em_i}\frac{\sqrt{\pi}}{2^{\chi+1}}\frac{\Gamma(\chi+1)}{\Gamma(\chi+5/2)}(2E_0)^{\chi+5/2}\nonumber\\
&{}&\times\Bigg[A\cos(\alpha{A}_x)\int_{0}^{2\pi}\int_{0}^{1}r(1-r^2)^{3/2+\chi}\cos(\alpha'{r}\cos\theta)drd\theta\nonumber\\
&{}&+\exp(\gamma{A_y})\int_{0}^{2\pi}\int_{0}^{1}r(1-r^2)^{3/2+\chi}\exp(\gamma'{r}\sin\theta)drd\theta+\frac{\pi{B}}{\chi+5/2}\Bigg],\nonumber
\end{eqnarray}
where, as in the previous two examples, $\alpha'$ and $\gamma'$ are given by Eqs. (\ref{a_prime}) and (\ref{b_prime}).
As in the previous two examples, the remaining integrations can be carried out by using the formulae in the Appendix. The pressure $P_{zz,g}$ is then given by
\begin{eqnarray}
P_{zz,g}(A_x,A_y)&=&(2\pi)^{3/2}\frac{f_0(m_e+m_i)e}{m_em_i}\Gamma(\chi+1)(2E_0)^{\chi+5/2}\frac{J_{\chi+5/2}(i\gamma')}{(i\gamma')^{\chi+5/2}}\nonumber\\
&{}&\times\Bigg[A\left(\frac{i\gamma'}{\alpha'}\right)^{\chi+5/2}\frac{J_{\chi+5/2}(\alpha')}{J_{\chi+5/2}(i\gamma')}\cos(\alpha{A_x})+\exp(\gamma{A_y})\nonumber\\
&{}&+\frac{B(i\gamma')^{\chi+5/2}}{2^{\chi+3/2}\Gamma(\chi+7/2)J_{\chi+5/2}(i\gamma')}\Bigg].\label{pzzzz}
\end{eqnarray}
The conditions (\ref{hhh})-(\ref{hhh2}) for the pressure (\ref{pzzzz}) to be equal to (\ref{p}) are then given by,
\begin{eqnarray}
\frac{\beta_e+\beta_i}{\beta_e\beta_i}n_0&=&(2\pi)^{3/2}\frac{f_0(m_e+m_i)e}{m_em_i}\Gamma(\chi+1)(2E_0)^{\chi+5/2}\label{ccc1}\nonumber\\
&{}&\times\frac{J_{\chi+5/2}(i\gamma')}{(i\gamma')^{\chi+5/2}}\\
\frac{1}{2}&=&A\left(\frac{i\gamma'}{\alpha'}\right)^{\chi+5/2}\frac{J_{\chi+5/2}(\alpha')}{J_{\chi+5/2}(i\gamma')}\label{ccc2}\\
b&=&\frac{B(i\gamma')^{\chi+5/2}}{2^{\chi+3/2}\Gamma(\chi+7/2)J_{\chi+5/2}(i\gamma')}\label{ccc3}.
\end{eqnarray}
Conditions (\ref{ccc1}) to (\ref{ccc3}) reduce to the two conditions
\begin{eqnarray}
\frac{J_{\chi+5/2}(\gamma')}{\gamma'^{\chi+5/2}}&>&0,\label{ccc4}\\
\frac{J_{\chi+5/2}(\alpha')}{\alpha'^{\chi+5/2}}&>&0\label{ccc5},
\end{eqnarray}
on removing known positive quantities from the inequalities and using the fact that $J_\nu(iz)=i^\nu{J}_\nu(z)$. The condition $B>A$ gives the following condition on $b$
\begin{eqnarray}
b>\frac{(\alpha')^{\chi+5/2}}{2^{\chi+5/2}\Gamma(\chi+7/2)J_{\chi+5/2}(\alpha')}.\nonumber
\end{eqnarray}
The condition (\ref{ccc4}) contains a root of $\gamma'$ meaning that, in general, $\gamma'$ must be positive, and so the conditions (\ref{ccc4}) and (\ref{ccc5}) are satisfied in the regions where $J_{\chi+5/2}(\alpha')$ and $J_{\chi+5/2}(\gamma')$ are positive (note again that $\alpha'=\sqrt{2E_0}\alpha>0$ through the definition (\ref{ind2})). Note that, when $\chi$ is an integer, the identity (\ref{spherical_bessel}) can be used to express the Bessel functions $J_{n+1/2}$ in terms of spherical Bessel functions $j_n$. This was done in the previous example (which corresponds to $\chi=0$). In that example, the conditions (\ref{ccc1}) to (\ref{ccc3}) were expressed in terms of $\sin\alpha'$, $\cos\alpha'$, $\sinh\gamma'$ and $\cosh\gamma'$ (from the spherical Bessel function $j_1$ given by (\ref{sph_bess1})) and it was observed that $\gamma'$ could be positive or negative without violating the conditions on the parameters (the fractional part of the power of $\gamma'$ cancelled out).

\section{Summary and Conclusions}

\label{sec:summary}

\noindent A method has been presented, which shows how to calculate a new family of distribution functions for the force-free Harris sheet, providing certain conditions on the parameters are satisfied. This family includes the known distribution function found by Harrison and Neukirch \cite{Harrison-2009b}, as well as distribution functions with a different dependence on the particle energy, but with the same dependence on the canonical momenta. Three specific examples have been given of possible solutions, to illustrate how the method can be used. In each example, the conditions on the parameters have been stated explicitly, and the validity of these conditions has also been discussed. So far, the force-free Harris sheet is the only magnetic field profile for which nonlinear force-free Vlasov-Maxwell equilibria are known. If an equilibrium solution was discovered for another field profile, then this method could potentially be used to extend the known solution to a whole family of solutions for that particular field profile. 

\acknowledgments{
\noindent The authors acknowledge financial support by the Leverhulme Trust and by the UK's Science and Technology Facilities Council.
}

%\bibliographystyle{unsrt}
%\bibliography{fiona}

  \renewcommand{\theequation}{A\arabic{equation}}
  % redefine the command that creates the equation no.
  \setcounter{equation}{0}  % reset counter 
  \section*{APPENDIX: EVALUATION OF INTEGRALS}

%\appendix

%\section{Evaluation of Integrals}

\noindent In each of the three examples in Section \ref{sec:examples}, after changing to a cylindrical coordinate system $(r,\theta)$, the $\theta$-integrations can be carried out by using the formulae
\begin{eqnarray}
\int_{0}^{2\pi}\cos(a\cos\theta)d\theta&=&2\pi{J_0}(a),\label{theta2}\\
\int_{0}^{2\pi}\exp(a\sin\theta)d\theta&=&2\pi{I_0}(a),\label{theta1}
\end{eqnarray}
where $J_0$ and $I_0$ are Bessel functions of the first and second kind, respectively. The $r$-integrals then take the following form,
\begin{eqnarray}
\int_{0}^{1}r(1-r^2)^{\lambda}J_0(ar)dr\label{j_int},\nonumber\\
\int_{0}^{1}r(1-r^2)^{\lambda}I_0(ar)dr\label{i_int}.\nonumber
\end{eqnarray}
Integrals of the form (\ref{j_int}) can be evaluated firstly by using the substitution $r=\sin{t}$, and then by using the formula
\begin{equation}
\int_{0}^{2\pi}J_{\mu}(z\sin{t})\sin^{\mu+1}t\cos^{2\nu+1}tdt=\frac{2^\nu\Gamma(\nu+1)}{z^{\nu+1}}J_{\mu+\nu+1}(z),\nonumber
\end{equation}
which is valid for $\Re\mu,\Re\nu>-1$. This gives
\begin{equation}
\int_{0}^{1}r(1-r^2)^\lambda{J}_0(ar)dr=\frac{2^\lambda\Gamma(\lambda+1)}{a^{\lambda+1}}J_{\lambda+1}(a)\label{app1}.
\end{equation}
Integrals of the form (\ref{i_int}) can be evaluated by firstly using the identity
\begin{equation}
I_0(a)=J_0(ia)\label{app2},\nonumber
\end{equation}
and then by using the same steps that were used to evaluate the integrals of the form (\ref{j_int}). This gives
\begin{equation}
\int_{0}^{1}r(1-r^2)^\lambda{I}_0(ar)dr=\frac{2^\lambda\Gamma(\lambda+1)}{(ia)^{\lambda+1}}J_{\lambda+1}(ia)\label{app3}.
\end{equation}

%\section{Spherical Bessel Functions}

%Bessel functions of the first kind $J_{n+1/2}$, which are of fractional order $n+1/2$, can be expressed in terms of spherical Bessel functions $j_n$ of integer order $n$, through the following identity
%\begin{equation}
%j_n(z)=\sqrt{\frac{\pi}{2z}}J_{n+1/2}(z).
%\end{equation}
%The first two spherical Bessel functions are given by
%\begin{eqnarray}
%j_1(z)&=&\frac{\sin{z}}{z^2}-\frac{\cos{z}}{z},\\
%j_2(z)&=&\left(\frac{3}{z^2}-\frac{1}{z}\right)-\frac{3}{z^2}\cos{z}.
%\end{eqnarray}

%\bibliographystyle{unsrt}
%\bibliography{fiona}

\end{document}